\documentclass{vldb}
\usepackage{graphicx}
\usepackage{balance}  % for  \balance command ON LAST PAGE  (only there!)
\usepackage{hyperref}
\usepackage{tikz}
\usepackage{blindtext}
\usepackage[final,layout={index}]{fixme}
\usepackage{enumitem}
\mathchardef\mhyphen="2D
\begin{document}
% ****************** TITLE ****************************************

\title{Needles in the `Sheet'stack: Augmented Analytics to get Insights from 
Spreadsheets}

\author{
\alignauthor
Medha Atre, Anand Deshpande, Reshma Godse, Pooja Deokar, Sandip Moharir
\and
\alignauthor
Dhruva Ray, Akshay Chitlangia, Trupti Phadnis, Yugansh Goyal\\
\affaddr{Persistent Systems}\\
\affaddr{Pune, India}\\
\email{firstname\_lastname@persistent.com}
}

\maketitle
    \begin{tikzpicture}[remember picture,overlay]
    \node[anchor=north west,inner sep=3pt]%
        at (current page.north west)
        {\includegraphics[height=2cm]{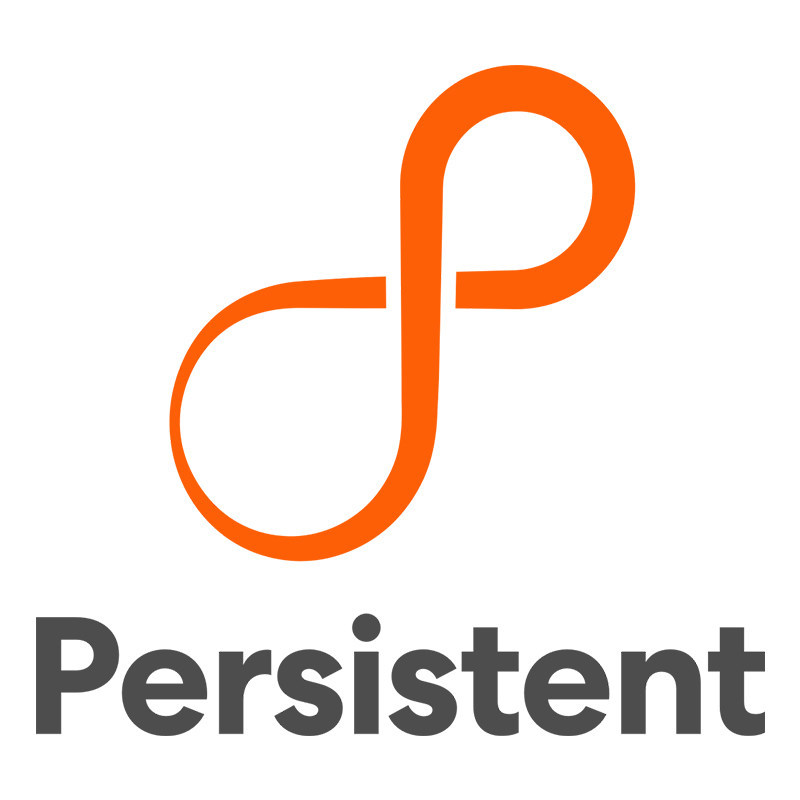}};
    \end{tikzpicture}

\begin{abstract}
Business intelligence (BI) tools for database analytics have come a long way 
and nowadays also provide ready insights or visual query explorations, e.g. 
QuickInsights by Microsoft Power BI, SpotIQ by ThoughtSpot, Zenvisage, etc.
In this demo, we focus on providing insights by examining periodic 
spreadsheets of different reports (aka views), without prior knowledge of 
the schema of the database or reports, or data information. Such a 
solution is targeted at users without the familiarity with the database 
schema or resources to conduct analytics in the contemporary way.
\end{abstract}

\section{Introduction} \label{sec:introduction}

Business Intelligence (BI) tools built on top of database systems provide 
excellent data analytics capabilities. E.g., trends of revenues of 
different companies over the past six months, or different departmental store 
products' sales comparison over the past three months etc. 
However, user needs a fair bit of knowledge of the underlying database 
schema and acquaintance with the BI dashboard. This can be 
particularly challenging for users and organizations that are not tech savvy or 
do not have time and resources to do so. A recent BI tools 
survey\footnote{\scriptsize{\url{https://bi-survey.com/bi-deployment}}} 
highlights that despite the market presence of a large variety of BI tools, 
about 38\% of the users continue to use spreadsheets as the main 
reporting and analytics tool. The use of BI tools remains low at an 
average of 10-15\% of the users. While spreadsheet tools like Microsoft 
Excel provide advanced analytics abilities, e.g., pivot 
tables, visual charts, ready-to-use functions etc, they still have the 
bottleneck that the user requires learning about these techniques and knowledge 
of the database and spreadsheet schema. Also it is our observation that 
spreadsheet reports are often generated targeting a wide variety of audiences 
within the organization making them bulky with tens of columns and thousands of 
rows.

\textit{Augmented analytics} proposes to take this \textit{learning} effort
off the user's shoulder, with the help of machine learning techniques for 
automated data cleaning, preparation, and insight 
discovery\footnote{\scriptsize{\url{https://www.gartner.com/en/documents/3773164
}}}. For example, if a user wants to find the sharpest sales trend (rising or 
falling) among thousands of departmental store products, an augmented analytics 
tool can provide such \textit{insights} automatically. Existing BI 
tools have indeed started moving in this direction, such as QuickInsights by 
Microsoft Power BI \cite{quickinsights17,quickinsights19}, SpotIQ by ThoughtSpot 
etc. These tools, in their present form, are tightly coupled with 
an existing BI dashboard interface and a backend database. However, as we noted 
above, a sizable percent of users still use spreadsheets as the main 
reporting and analytics tool.

Our main contributions through this demo proposal are:
\begin{enumerate}[leftmargin=1.5em,noitemsep,nosep]
 \item Finding \textit{insights} from a series of spreadsheets of a report, 
without requiring pre-training or prior knowledge of the spreadsheet schema.
 \item Allowing users to interact with the systems using a chatbot and 
semi-structured English commands to set individual preferences, and give 
\textit{personalized insights} from the same spreadsheets there onward for 
different users.
\end{enumerate}
For instance, CEO and marketing manager of a company can both get different 
personalized insights from the same series of spreadsheets. Our goal is to 
build a system similar to the modern social media, where different newsfeeds 
are generated for different users from the same underlying data. Our system's 
focus is on the personalized newsfeed of insights generated from the 
\textit{same} spreadsheets of organization reports, consumed over emails, 
chatbots, RSS etc. 

\section{Architecture}\label{sec:architecture}

\begin{figure}
 \includegraphics[scale=0.55]{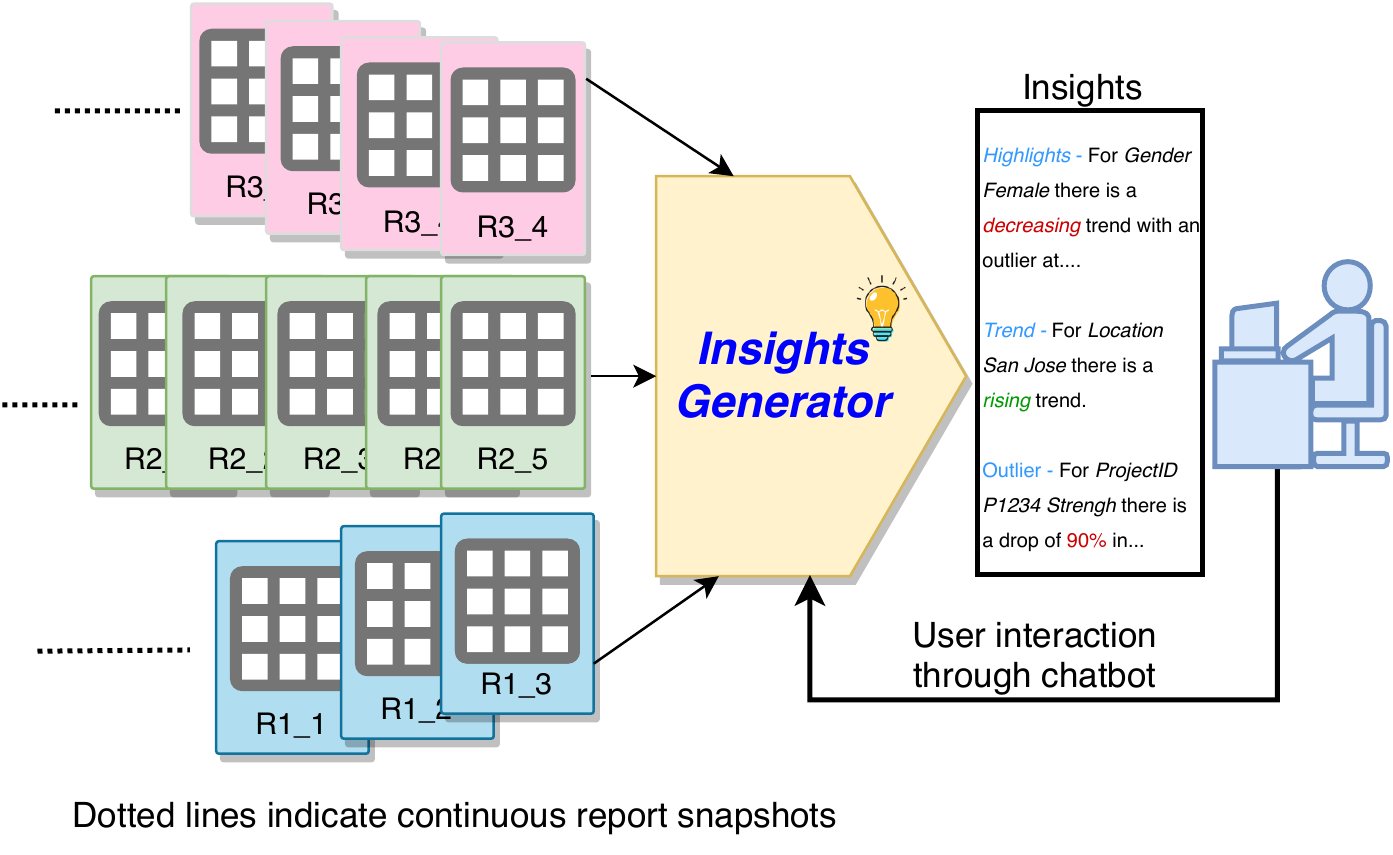}
 \caption{Architecture} \label{fig:arch}
\end{figure}

Figure \ref{fig:arch} shows the high level architecture of our system. 
\textit{Insights Generator} runs continuously accepting spreadsheets of 
different reports. In Figure \ref{fig:arch}, $R1, R2, R3$ are 
three different types of reports, e.g., $R1$ can be product sales report, 
$R2$ new joinee and attrition report, and $R3$ bug report. $R1_1, R1_2...$ 
represent periodic spreadsheets of report type $R1$ and so on. 
The engine does not impose any limit on the types of reports, the number of 
spreadsheets of each report, or their periodicity. It treats each report type 
and its spreadsheets independently.

The insight generator has three main components -- (1) backend analysis unit, 
(2) frontend for insights delivery and user interaction through chatbot, (3) a 
middle layer for communicating backend insights in JSON format to the frontend. 
Frontend in turn consists of two subcomponents -- (a) an insight delivery 
mechanism (currently through email), and (b) user interaction and 
personalization through chatbot using semi-structured English language commands. 
Note that through this 
personalization we enable different users to get different insights in the 
successive spreadsheets of the same report type.  We have deployed components of 
our engine using Microsoft Azure framework, but they can be deployed on any 
other suitable cloud or independent architecture.

\subsection{Data Cleaning and Preparation} \label{sec:dataprep}

Our insight generator consumes reports in the spreadsheet format, which may not 
be strictly structured, and may have free text in the top rows or leftmost 
columns. We extract the \textit{most significant table} from such sheets as 
follows -- from the top of the sheet, designate the first row with maximum 
non-empty columns as the header, discard any leftmost empty columns from 
this header, and extract the table under.

Without losing generality, let the spreadsheets of a report type $R1$ have 
columns $c_1, c_2, c_3...$. We categorize these into four mutually exclusive 
sets -- (1) primary key attributes (K), (2) categorical attributes (C), (3) 
numeric attributes (N), and (4) all others. Presently we disregard \#4 all 
other attributes. Due to space constraints, we briefly describe our method of 
identifying these four attribute types using heuristic measures.

For primary key attributes (K), we assume that typically they appear in the 
first five columns of a spreadsheet, and identify them by considering the total 
number of unique values in these columns and the total number of rows in the 
spreadsheet. In case if no such attribute or combination of them exist, we keep 
primary key attributes empty. We identify categorical attributes (C) using 
heuristic measures of the \textit{data-type} of the column and the 
ratio of total non-empty rows in that column to the unique values in it 
($\#nonemptyrows/ \#uniq val$). If the spreadsheet has more than five 
categorical attributes we sort all the categorical attributes in the descending 
order of ($\#nonemptyrows/ \#uniq val$), and uniformly pick five 
attributes from this sorted order. From these five categorical attributes, we 
create ${5 \choose 2} + {5 \choose 1} = 15$ attribute combinations for analysis. 
Note that our engine allows end user to change this choice or combination for 
future analysis as elaborated later. We consider all the numeric attributes (N) 
without discarding any.

\textbf{Timeseries:}
% We generate insights on at least five or more spreadsheets 
% of a report, by creating three types of \textit{time-series} from them as 
% follows. Treatment of less than five spreadsheets is given later.
Let us consider $R1_1, R1_2...R1_s$ spreadsheets of report $R1$, 
where $R1_1$ has the oldest timestamp and $R1_s$ the latest. We first choose 
$R1_1$, and consider each primary key (composite) value $k_i$ of $K$ 
attributes. This is a composite key if $K$ has more than one attribute. We 
consider each numeric attribute $n_j \in N$, and create a timeseries key $(k_i, 
n_j)$ (key-attribute KA pair). E.g., if $R1$ has \textit{Product-ID} as the 
primary key, with \textit{Sales} as a numeric attribute, the timeseries key for 
product with ID $P1234$ is \textit{(P1234, Sales)}. Let the value of 
\textit{Sales} for 
$P1234$ in $R1_1$ be $y_1$, in $R1_2$ be $y_2$ and so on. Thus from 
$R1_1...R1_s$ spreadsheets, we form a time series for key \textit{(P1234, 
Sales)} $\rightarrow [(t_1, y_1), (t_2, y_2)...(t_s, y_s)]$, where $t_1...t_s$ 
are the timestamps of the spreadsheets. A spreadsheet, say $R1_4$, may not have 
the specific key $P1234$, in which case we do not enter $(t_4, y_4)$ value in 
the timeseries, thus accommodating for disappearing and reappearing entities. 
This is done for each unique primary key value and unique numeric attribute in 
all the spreadsheets. We call these timeseries numeric attribute timeseries or 
\textbf{NTS}.

Next, for each spreadsheet $R1_i$, we consider each numeric column $n_j \in N$, 
and order $n_j$'s values corresponding to each primary key value and add 
a numeric column of these ranks, we call this $\mathit{n_j\mhyphen rank}$. 
E.g., within the \textit{Sales} column in $R1_1$, let $P1234$ be 
\textit{\$1000}, $P2345$ be \textit{\$500}, and $P3456$ be \textit{\$1200}, then 
the relative ascending order of $P1234, P2345, P3456$ for \textit{Sales} column 
is $2, 1, 3$ and we add a column \textit{Sales-rank}. Using the same procedure 
described above for NTS timeseries, we create rank timeseries \textbf{RTS} for 
each hybrid timeseries key $(k_i, \mathit{n_j\mhyphen rank}) \rightarrow [(t_1, 
r_1), (t_2, r_2)...(t_s, r_s)]$, where $r_1...r_s$ are the values corresponding 
to $k_i$ in $\mathit{n_j\mhyphen rank}$ column of each spreadsheet. In all we 
get $(2 * |\bigcup_{R1_1...R1_s} uniq(K) \times N|)$ NTS and RTS timeseries 
over all the spreadsheets\footnote{\scriptsize{If spreadsheets do not have any 
primary keys, we only process categorical attributes discussed next.}}.

Recall that we pick five categorical attributes, if more than five are present, 
and prepare maximum fifteen combinations of them. For each spreadsheet, we 
consider each $c_k$  categorical attribute combination of these maximum $15$ 
combinations. We compute the total number of rows $u$ for a unique value $v_l 
\in c_k$. This is a composite value if $c_k$ has two categorical attributes. We 
do this for each spreadsheet, and form a timeseries for $(v_l, c_k) \rightarrow 
[(t_1, u_1), (t_2, u_2)...(t_s, u_s)]$ for each $(v_l, c_k)$ value-attribute 
(VA) pair across all the spreadsheets. We call these categorical attribute 
timeseries or \textbf{CTS}. Thus given timestamp ordered spreadsheets of a 
particular report, we form three types of timeseries -- NTS, RTS, CTS.

\subsection{Analytics} \label{sec:analytics}
For the analytical processing, we consider only timeseries that have more than 
five points in them. Shorter timeseries are discussed after this under 
``\textit{LT5 Mean, Variance}''.

\textbf{Linear Regression:}
Recall that our engine neither assumes any information about the spreadsheet 
schema and data domain, nor is it pre-trained on any existing corpus of schemas 
such as \url{schema.org}. We use unsupervised learning methods based on -- 
(1) trend analysis (linear regression), (2) mean squared error (MSE) of the 
fitted trend line, and (3) outlier scores of data points within the fitted 
trend, i.e., Cook's 
Distance\footnote{\scriptsize{\url{
https://en.wikipedia.org/wiki/Cook's\_distance}}}. As given in Section 
\ref{sec:dataprep}, we form NTS, RTS, CTS consisting of several timeseries 
based on the unique values of primary 
keys and categorical attributes in the spreadsheets. For each timeseries in 
RTS, NTS, CTS, we do linear regression and fit a trend line. We compute 
the mean squared error (MSE) of this fit, and compute Cook's Distance of each 
data point $(t_i, y_i)$, $(t_i, r_i)$, and $(t_i, u_i)$ in each time series. 
For each time series, we pick the data point with highest Cook's Distance. Thus 
at the end of this exercise, each timeseries in NTS, RTS, and CTS has a (1) 
line equation with slope $m$ and intercept $b$, (2) MSE $mse$, and (3) a point 
with maximum Cook's Distance $mcd$\footnote{\scriptsize{Such a Cook's Distance 
always tends to order time series elements with large numbers first. But this 
can be configured to consider \textit{normalized} numbers, for a fair treatment 
to the smaller numbers.}}. We use these three features for deciding 
relative ranking of timeseries within each of NTS, RTS, CTS as follows. 
The procedure is applied same way for each of NTS, RTS, CTS group timeseries 
and hence we do not mention group names explicitly. We sort timeseries in the -- 
(1) descending order of $m^2$ (sharpness of the slope irrespective of whether 
the slope is rising or falling), (2) descending order of $mse$, and (3) 
descending order of $mcd$. In all, we get \textit{nine} orderings of timeseries, 
three each for NTS, RTS, CTS..

\textbf{LT5 Mean, Variance:}
When the length of a timeseries is less than or equal to five, we only compute 
the mean and variance of the timeseries points, for each timeseries in NTS, 
RTS, CTS. Thus for each shorter timeseries in NTS, RTS, CTS we compute -- (1) 
mean $\mu$, and (2) variance $\alpha^2$.

\textbf{Diff of the latest two:} We consider only the latest two spreadsheets of 
a report type, and compute the difference between them. This is 
achieved by using timeseries of NTS, RTS, CTS, and considering the last two 
points within them, if their timestamps correspond to the latest two reports. 
E.g., if a timeseries within NTS group is $(k_i, n_j) \rightarrow (t_1, 
y_1)...(t_9, y_9), (t_{10}, y_{10})$,then we compute $(y_{10} - y_9)^2$ for the 
given $(k_i, n_j)$ value, if $t_{10}$ is of the latest spreadsheet. We repeat 
the same process for RTS and CTS timeseries.

\textbf{New Entities and Attributes:} Considering only the latest spreadsheet 
of a report type, say $R1$, we compute if there are any new primary key values 
or attributes added to it. In our observation, the report schema can undergo 
slight changes over time. Our procedure of timeseries composition outlined 
above can accommodate such changes, because we form timeseries for each 
unique key-attribute (KA) or value-attribute (VA) pair. From these various 
metrics computed on timeseries and spreadsheets, we compute \textit{insights} 
(the most significant observations) as described in Section \ref{sec:insights}.

\subsection{Insights} \label{sec:insights}
We generate insights in four categories -- (1) Overall highlight, 
(2) most significant sharp and flat trends, (3) most significant outlier, (4) 
most significant difference of the latest two spreadsheets (Delta).

\textbf{Highlight}: Recall from Section \ref{sec:analytics} that after linear 
regression we sort timeseries by $m^2$, $mse$, and $mcd$ to get three sorted 
orders of timeseries in NTS, RTS, CTS each. Within the NTS timeseries, we 
compute a \textit{composite sorted order} of each timeseries by multiplying 
sorted indices of it in each of the three sort orders, i.e., for the timeseries 
of $(k_i, n_j)$, its composite rank is $o_{m^2} * o_{mse} * o_{mcd}$, where 
$o_{m^2},o_{mse}, o_{mcd}$ are indices of $(k_i, n_j)$ in the sorted orders 
of $m^2, mse, mcd$ respectively. From the composite rank we pick the top 
one. In case of a tie, we pick one randomly. We repeat this process for RTS, 
CTS group of timeseries too, and pick the timeseries with the top composite 
rank. Intuitively, the highlighted insight from NTS, RTS, CTS are timeseries 
that have sharp trend (rising or falling), have high fluctuations, and an 
outlier with relatively higher residual error.

\textbf{Trend:} Disregarding the timeseries picked in the Highlight (to avoid 
redundancy), next we pick the top timeseries from the $m^2$ sorted order for 
each of the NTS, RTS, CTS group. Additionally we also pick the last timeseries 
from $m^2$ order. Thus the Trend insights have sharp rising or falling and the 
flattest trends.

\textbf{Outlier:} Next, disregarding the timeseries picked in Highlight and 
Trend, we pick the top timeseries from the sorted order of $mcd$ (Cook's 
Distance) for NTS, RTS, CTS each.

\textbf{Delta:} Considering the `\textit{Diff of the latest two}' as given in 
Section \ref{sec:analytics}, we pick the timeseries that shows the maximum 
change (Delta) in the latest two reports.

Note that we compute relative order of \textit{all} the timeseries formed over 
\textit{all} the numeric and selected categorical attributes (ref. 
Section \ref{sec:analytics}), and pick the top ones for insights. 
However, different users might be interested in focusing on different 
attributes. For this we provide a user interaction interface for 
\textit{personalization} as given in Section \ref{sec:userinterface}.

\begin{figure*}[ht]
 \centering
 \begin{minipage}{0.48\textwidth}
 \centering
 \includegraphics[height=2.5in, width=3.35in]{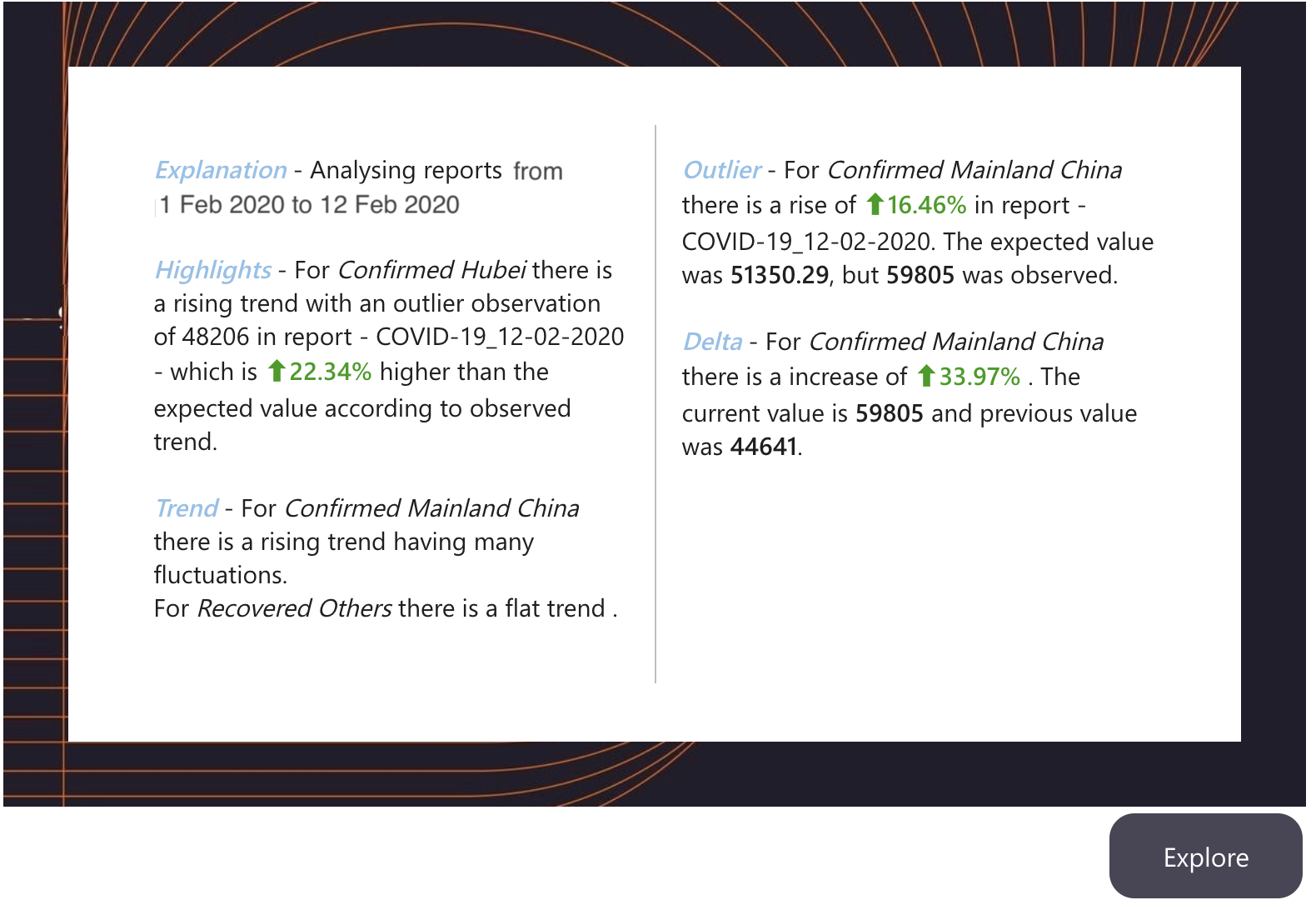}
 \caption{COVID-19 Insights 1--12 Feb} \label{fig:adcard1}  
 \end{minipage}
 \hfill
 \begin{minipage}{0.48\textwidth}
 \centering
 \includegraphics[height=2.5in, width=3.35in]{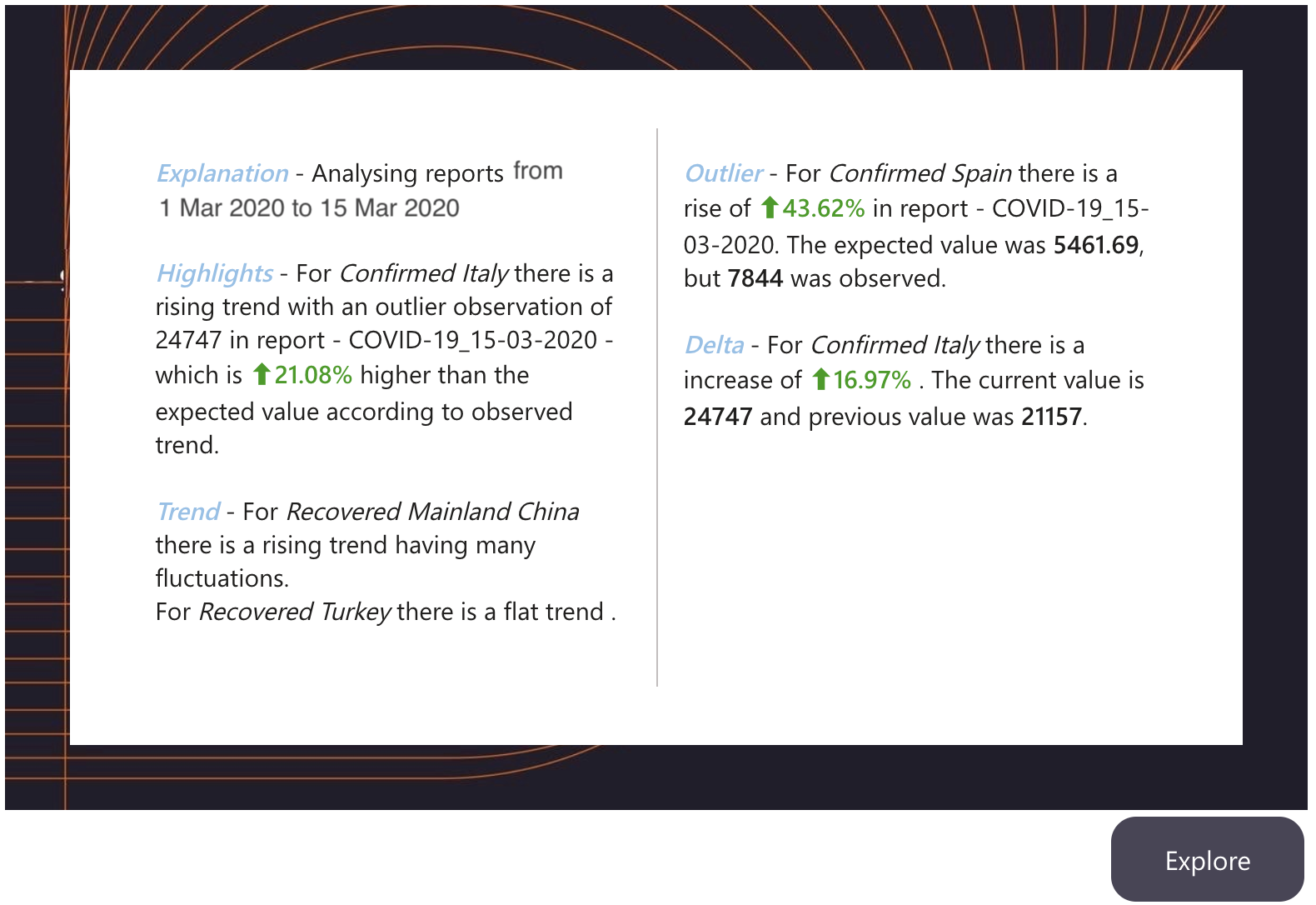}
 \caption{COVID-19 Insights 1--15 Mar} \label{fig:adcard2}
 \end{minipage}
\end{figure*}

\subsection{User Interaction} \label{sec:userinterface}
In spreadsheets with tens of columns and thousands of rows, the number of 
timeseries can run in several thousands, e.g., when we tested with Stock 
Exchange archive spreadsheets \cite{nse}, each spreadsheet has 
several numeric attributes and a couple of thousand rows, making around 20,000 
timeseries. Since our engine does not assume any prior knowledge of 
schema or data domain, we treat all of them at par and choose the top 
insights. But \textit{user-1} may be interested in insights from specific 
attributes, which may get overshadowed in the ranking from other insights.
Hence we let user interact with the engine and set personalized configuration 
and preferences using chatbot. Through this, for the next round of insights, 
each user will get insights according to their previously set preferences.
% Also we are working on providing a way of user to iteratively fetch more 
% insights from the ordered list at the backend in addition to the top ones 
% delivered to the user at first.

\subsection{Moving Window}
Recall from Figure \ref{fig:arch} that our engine can continuously 
churn periodically arriving spreadsheets, and generate a fresh set of insights 
after every new spreadsheet. We surmise that a typical user is most interested 
in finding out insights from the latest few spreadsheets. Thus we provide a 
configurable option to have a \textit{moving window} over spreadsheets sorted 
by their timestamp. E.g., for report $R1$, a user wants to consider only the 
latest 10 spreadsheets for insight generation. Thus when a new spreadsheet of 
$R1$ arrives, we move the window by discarding the $10^{th}$ \textit{oldest} 
spreadsheet and adding the latest in the window.

\section{Setup} \label{sec:setup}
Our engine's three main components (1) analysis or spreadsheet processing, (2) 
frontend for insight delivery and user feedback, (3) middle layer for 
communicating insights from analysis to the frontend using JSON, are 
implemented using Python 3.0. Currently our setup is deployed using 
Microsoft Azure infrastructure and works as follows.

We assume that a spreadsheet of a report is sent to several people in the 
organization over email. In addition to these people, the spreadsheet is also 
sent to an email robot listening service for our engine, e.g., 
\textit{insightalias@persistent.com}. Treating the subject of the email 
as the report type, the service creates a separate storage for each report, 
runs the procedure given in sections \ref{sec:dataprep}, \ref{sec:analytics}, 
and \ref{sec:insights}, and sends the first set of insights to all the 
recipients of the report over an email, without any inputs or 
configuration asked from the user. In Figure \ref{fig:adcard1} and 
\ref{fig:adcard2}, we have shown example of such insights generated on COVID-19 
data acquired from \cite{covidva}. Figure \ref{fig:adcard1} shows top insights 
generated on the spreadsheets between February 1--12 and Figure 
\ref{fig:adcard2} shows insights for March 1--15. Comparing the two, we can 
note that for February, all the top insights came from the Chinese 
provinces, whereas by March when COVID-19 spread in other countries, and 
its spread in China started dampening (due to the lock-down), the top insights 
started coming from Italy and Spain\footnote{\scriptsize{Presently we do not 
describe insights from RTS timeseries, that work is in progress.}}. 
\textit{\textbf{Note:} these insights were generated by our engine 
\underline{without} prior knowledge of the data, schema, or user inputs.}

Clicking the `\textit{Explore}' button opens a Microsoft Teams chatbot. Users 
can interact with the engine using simple semi-structured English commands,
personalize the configuration by choosing only some attributes according to 
their preference, and instantaneously get updated insights. The user can 
save this personal configuration. When the new spreadsheet arrives, the 
next insights are personalized using individual user's configuration.

\section{Related Work}

Microsoft Power BI's QuickInsights \cite{quickinsights17, quickinsights19}, 
ThoughtSpot's SpotIQ \cite{spotiq}, and other augmented analytics tools offered 
by the BI tools come closest to our system. However, all those systems are 
integrated with an existing BI tool, and thus require access to the customer 
database. In comparison, our tool is \textit{lightweight} which can work with 
spreadsheets without access to the full database or its schema. SeeDB 
\cite{seedb}, Zenvisage \cite{zenvisage16}, and their successor systems 
\cite{shapesearch} mainly focus on insightful visualizations of the underlying 
data. While their focus is on insightful visualizations, our focus is on 
insights presented in the English language. These systems expect knowledge of 
the underlying database schema to be able to identify the \textit{dimension} and 
\textit{measure} attributes. Since our system is targeted at users without the 
knowledge or resources to provide this information, we use heuristics to 
identify these attributes in the spreadsheets.

\section{Future Work}
In the ongoing enhancements, we are working on providing insights into 
identifying 
attribute correlations, to be able to provide insights into \textit{correlated 
outliers}, e.g., a sudden fall or rise in the sales of particular products due 
to the change in the store staff attendance (e.g., COVID-19 pandemic or weather 
patterns). Attributes which are dependent on other attributes will have similar 
timeseries, and generate redundant insights. Similar timeseries can be 
identified using a variety of techniques such as Earth Mover's Distance, 
Euclidean Distance, Dynamic Time Warp etc., and they can be clustered 
so as to efficiently provide \textit{unique} insights. We plan to analyse 
\textit{semantics} of attributes using NLP techniques, so as to make better 
decisions in deciding how to process them. Finally, while we keep the focus of 
the insights on the latest moving window, we intend to give a historical 
perspective for a particular timeseries.

\balance

\bibliographystyle{abbrv}
\scriptsize{
\bibliography{references}
}

\listoffixmes

\end{document}